\documentclass[aps,prd,floatfix,superscriptaddress,showpacs,showkeys]{revtex4}
\usepackage{amssymb}
\usepackage{amsmath}
\usepackage{amsfonts}
\usepackage{epsfig}
\usepackage{graphicx}
\usepackage{tabularx}
\usepackage{latexsym}
\usepackage{graphics}
\usepackage{verbatim}
\usepackage{color}
\newcommand{\seq}{\begin{subequations}}
\newcommand{\sen}{\end{subequations}}
\newcommand{\eq}{\begin{eqnarray}}
\newcommand{\en}{\end{eqnarray}}
\newcommand{\ra}{\rangle}

\def\dppm{D^\pm}
\def\dpp{D^+}
\def\dpm{D^-}
\def\dpmps{D^{\ast \mp}}
\def\dpps{D^{\ast +}}
\def\dpms{D^{\ast -}}

\def\d{D^0}
\def\db{\bar{D}^{0}}
\def\ds{D^{\ast  0}}

\def\dbs{\bar{D}^{\ast 0}}

\def\L2{\Lambda^2}

\def\eq{\begin{eqnarray}}
\def\en{\end{eqnarray}}

\def\dppm{D^\pm}
\def\dpp{D^+}
\def\dpm{D^-}
\def\dpmps{D^{\ast \mp}}
\def\dpps{D^{\ast +}}
\def\dpms{D^{\ast -}}

\def\d{D^0}
\def\db{\bar{D}^{0}}
\def\ds{D^{\ast  0}}

\def\dbs{\bar{D}^{\ast 0}}

\def\L2{\Lambda^2}

\def\eq{\begin{eqnarray}}
\def\en{\end{eqnarray}}

\def\L{{\cal L}}

\begin{document}

\title{\boldmath Radiative decay $Y(4260) \to X(3872) + \gamma$
involving\\
hadronic molecular and charmonium components}

\author{Yubing Dong}
\affiliation{Institute of High Energy Physics, Beijing 100049, 
People's Republic of China}
\affiliation{Theoretical Physics Center for Science
Facilities (TPCSF), CAS, Beijing 100049, People's Republic of China}
\author{Amand Faessler}
\affiliation{Institut f\"ur Theoretische Physik,  Universit\"at T\"ubingen,
Kepler Center for Astro and Particle Physics, \\
Auf der Morgenstelle 14, D--72076 T\"ubingen, Germany}
\author{Thomas Gutsche}
\affiliation{Institut f\"ur Theoretische Physik,  Universit\"at T\"ubingen,
Kepler Center for Astro and Particle Physics, \\
Auf der Morgenstelle 14, D--72076 T\"ubingen, Germany}
\author{Valery E. Lyubovitskij}
\affiliation{Institut f\"ur Theoretische Physik,  Universit\"at T\"ubingen,
Kepler Center for Astro and Particle Physics, \\
Auf der Morgenstelle 14, D--72076 T\"ubingen, Germany}
\affiliation{Department of Physics, Tomsk State University, 634050
Tomsk, Russia}
\affiliation{Mathematical Physics Department, 
Tomsk Polytechnic University, Lenin Avenue 30, 
634050 Tomsk, Russia} 

\date{\today}
\begin{abstract}
We apply a phenomenological Lagrangian approach to the
radiative decay $Y(4260) \to X(3872) + \gamma$. 
The $Y(4260)$ and $X(3872)$ resonances are considered as 
composite states containing both molecular hadronic and 
charmonium components. Having a leading molecular
component in the $X(3872)$ and a sole molecular configuration
for the $Y(4260)$ results in a prediction compatible with present data. 
\end{abstract}

\maketitle

\section{Introduction}

The study of the radiative decay $Y(4260) \to X(3872) + \gamma$ among
other production and decay modes can
give additional information on the nature of the exotic states
$Y(4260)$ and $X(3872)$. 

In Refs.~\cite{Faessler:2007gv}-\cite{Dong:2013kta} 
we proposed and developed a phenomenological Lagrangian approach for the study
of exotic mesons and baryons as hadronic molecules with a possible 
admixture of quark-antiquark (in case of mesons) 
and three-quark (in case of baryons) components. 
To set up the bound state structure of new exotic states we used 
the compositeness condition~\cite{Weinberg:1962hj}-\cite{Anikin:1995cf} 
which is the key ingredient of our approach. 
In Refs.~\cite{Efimov:1993ei,Anikin:1995cf}
and~\cite{Faessler:2007gv}-\cite{Dong:2012hc} it was proved that
this condition is an important and successful quantum field theory
tool for the study of hadrons and exotic states as bound states of
their constituents. 
In particular, in Refs.~\cite{Dong:2008gb,%
Dong:2013iqa,Dong:2013kta} we considered strong and radiative decays
of the $X(3872)$ and $Y(4260)$. 
A detailed analysis of the $X(3872)$ decay modes in the 
picture, where this state is considered as a superposition of 
the molecular $D^0D^{\ast 0}$, $D^\pm D^{\ast \mp}$, $J/\psi \omega$, 
$J/\psi \rho$ and $c\bar c$ charmonium configurations,  
has been considered in Ref.~\cite{Dong:2009uf}. It was shown that 
a compact charmonium component plays an important role in radiative 
transitions $X(3872) \to J/\psi(\psi(2s)) \gamma$, which was 
in agreement with predictions obtained in potential 
models~\cite{Barnes:2003vb}-\cite{Li:2009zu}.  

Here we present a quantitative study of the
radiative decay mode $Y(4260) \to X(3872) + \gamma$ (see also the
discussion in Refs.~\cite{Guo:2013zbw,Ma:2014ofa}) by using the
phenomenological Lagrangian approach developed previously in
Refs.~\cite{Faessler:2007gv}-\cite{Dong:2009tg}. 
We assume that both $Y(4260)$ and $X(3872)$ states are composite states 
containing both extended hadronic molecular (with a typical size of 
$r > 1.5-2$ fm) and compact charmonium (with a typical size $r < 1$ fm) 
components. In Ref.~\cite{Dong:2009uf} such a scenario was considered 
for the $X(3872)$ state, and here we extend this idea to the case 
of the $Y(4260)$ state.  
Following Refs.~\cite{Zhu:2005hp,Ding:2008gr,Wang:2013cya,%
Guo:2013zbw,Dong:2013kta}, 
where the molecular $DD_1(2420)$ 
assignment for the $Y(4260)$ state was proposed 
and studied, we assume that the $Y(4260)$ is an composite isosinglet 
state containing a molecular component made up of the
pseudoscalar $D(1870)$ and the axial $D_1(2400)$ charm mesons 
and $c\bar c$ charmonium configuration 
\eq
|Y(4260)\ra \ = \ 
\frac{1}{2} \, \Big| \bar{D}_1 D - \bar{D} D_1
\Big\ra \, \cos\theta_Y + |c \bar c \ra \, \sin\theta_Y  
\en
with $J^P=1^-$. $\theta_Y$ is the mixing angle between the hadronic 
and the charmonium components. Here  and in the following we will use
the notation $D = (D^+, D^0)$, $D^\ast = (D^{\ast\,+}, D^{\ast\,0})$, 
$D_1 = (D_1^+, D_1^0)$ which stand for the doublets of pseudoscalar, 
vector and axial-vector $D$ mesons. 
In this paper we use the convention that
the pseudoscalar and axial mesons do not change the sign under
charge-parity transformation while the vector meson changes the
sign. In previous papers we used another convention, but it does not
affect the results.

The $X(3872)$ with $J^P = 1^+$ is set up as a superposition 
of the molecular $D^0 \bar D^{\ast\,0} - \bar D^0 D^{\ast\,0}$,
$D^+ D^{\ast\,-} - D^- D^{\ast\,+}$,
$J/\psi \omega$, $J/\psi \rho$, and the $c \bar c$ components 
as proposed in Ref.~\cite{Swanson:2003tb} and developed in 
Ref.~\cite{Dong:2009uf}
\eq
|X(3872)\ra &=& \cos\theta_X\Bigg [
\frac{Z_{\d\ds}^{1/2}}{\sqrt{2}}( |\d\dbs \ra - |\ds\db \ra )
+ \frac{Z_{\dppm\dpmps}^{1/2}}{\sqrt{2}}( | \dpp\dpms \ra
- | \dpm\dpps \ra )  \nonumber \\
&+& Z_{J/\psi\omega}^{1/2} | J/\psi \omega \ra
 + Z_{J/\psi\rho}^{1/2} | J/\psi \rho \ra \Bigg ]
 + \sin\theta_X \, c\bar{c} 
\,.
\en
The mass of the $X(3872)$ state ($M_X$) is expressed in terms of the masses of 
the constituents $D^0$ and $D^{\ast 0}$ and the binding energy $\epsilon_{X}$
\eq
M_{X}=M_{D^0}+M_{D^{\ast\,0}}-\epsilon_{X}\,,
\en
where $\epsilon_{X}$ varies from 0 to 0.3 MeV. For these values of 
$\epsilon_{X}$ the charged combination $D^+ D^{\ast\,-}  - D^- D^{\ast\,+}$ 
and the other two components $J/\psi\rho$ and $J/\psi\omega$ 
are significantly suppressed~\cite{Swanson:2003tb,Dong:2009uf}.
In particular, the probability for the leading molecular component 
$D^0 \bar D^{\ast\,0} - \bar D^0 D^{\ast\,0}$ 
with respect to the other hadronic components varies from 100\% to 92\%. 
The probabilities of the other hadronic components vary for 
the charged $D^+ \bar D^{\ast\,+} - D^- D^{\ast\,-}$ component 
from 0\% to 3.3\%, for the $J/\psi \omega$ component from 0\% to 4.1\% and 
for the $J/\psi \rho$ component from 0\% to 0.6\%. It should be clear that 
the $J/\psi \omega$ and $J/\psi \rho$ components do not contribute 
to the radiative decay $Y(4260) \to X(3872) + \gamma$, while 
the $D^+ \bar D^{\ast\,+} - D^- D^{\ast\,-}$ component could. However, 
the last contribution is further reduced due to the 
suppression of the coupling $g_{_{D^{\ast \, \pm } D^\pm_1 \gamma}}$ 
by a factor 4 in comparison to 
$g_{_{D^{\ast \, 0} D^0_1 \gamma}}$ (see details in Sec.II). 
For the mass of the $Y(4260)$ state we use the central value 
$M_Y = 4250$ MeV. 
 
As in the case of the $Y(4260)$ state 
we introduce a corresponding mixing angle $\theta_X$ encoding the 
mixing between the hadronic molecular and charmonium components 
in $X(3872)$.  In this paper we use the convention that
the pseudoscalar and axial mesons do not change the sign under
charge-parity transformation while the vector meson changes the
sign. In previous papers we used another convention, but it does not 
affect the results. 

The paper is organized as follows. In Sec.~II we briefly review the
basic ideas of our approach and show the effective Lagrangians for
our calculation. Then, in Sec.~III, we proceed to derive the width
of the radiative two-body decay mode $Y(4260)\to X(3872) + \gamma$.
In our analysis we approximately take into account the mass
distribution of the $Y(4260)$ state. Finally we present our
numerical results and compare to recent limits set by experiment.

\section{Basic model ingredients}

Our approach to the possible composite structure of $X(3872)$ and
$Y(4260)$ as mixed bound states containing respective hadronic molecular 
and charmonium components is based on interaction Lagrangians. They describe
the coupling of the respective states to their constituents
\eq\label{Lagr_general} 
{\cal L}_{H}(x) = H^{\mu}(x) \, 
\Big(\cos\theta_H g_H^M J_{H;\mu}^M(x) \,
+\, \sin\theta_H g_H^C J_{H;\mu}^C(x)\Big)  
\en 
where $H = X, Y$,  $g_{_{H}}^M$ and  $g_{_{H}}^C$ are the 
dimensionless couplings of the $H$ state to the molecular and charmonium 
components, respectively. Here, $J_H^M$ and $J_H^C$ are the respective 
interpolating hadronic and quark currents with quantum numbers 
of the $H$ state. In particular, these currents are written as
\eq  
J_{X;\mu}^M(x) &=& \frac{iM_X}{\sqrt{2}} 
\, \int d^4y \, \Phi_{X}^M(y^2) \bar D_{\mu}^{\ast 0}(x+y/2)
D^0(x-y/2)\, + \, {\rm H.c.}\,, \nonumber\\
J_{Y;\mu}^M(x) &=& \frac{iM_Y}{2}\, M_{Y} 
\, \int d^4y \, \Phi_{Y}^M(y^2) \, \bar{D}_{1\,\mu}(x+y/2) \, D(x-y/2) 
\,+\, {\rm H.c.}\,, \nonumber\\
& &\\
J_{X;\mu}^C(x) &=& 
\int d^4y \Phi_{X}^C(y^2)\bar{c}(x+y/2) \gamma_{\mu}\gamma^5 c(x-y/2)\,,
\nonumber\\
J_{Y;\mu}^C(x) &=& 
\int d^4y \Phi_{Y}^C(y^2)\bar{c}(x+y/2) \gamma_{\mu} c(x-y/2)\,,
\nonumber
\en
where $y$ is a relative Jacobi coordinate; 
$\Phi_H^M(y^2)$ and $\Phi_H^C(y^2)$ are 
correlation functions which describe the distribution of the
constituent mesons and charm quarks in the bound states $X(3872)$ 
and $Y(4260)$, respectively. 

A basic requirement for the choice of an explicit form of the
correlation function $\Phi_H(y^2)$ is that its Fourier transform
vanishes sufficiently fast in the ultraviolet region of Euclidean
space to render the Feynman diagrams ultraviolet finite. For
simplicity we adopt a Gaussian form for the correlation function.
The Fourier transform of this vertex function is given by
\eq
\label{corr_fun} \tilde\Phi_H^I(p_E^2/(\Lambda_H^I)^2) \doteq \exp( -
p_E^2/(\Lambda_H^I)^2)\,, \quad I = M, C 
\en
where $p_{E}$ is the Euclidean Jacobi momentum. $\Lambda_H^I$ stand
for the size parameters characterizing the distribution of the two
constituent mesons $(I=M)$ or constituent charm quarks $(I=C)$ 
in the $X(3872)$ and $Y(4260)$ systems. Note that
these scale parameters have been constrained before as 
$\Lambda_X = 0.5$ GeV~\cite{Dong:2009uf}, 
$\Lambda_Y = 0.75$ GeV~\cite{Dong:2013kta} 
and $\Lambda_H^C = 3.5$ GeV~\cite{Dong:2009uf}. 
The respective coupling constants $g_{_H}^I$ are
determined by the compositeness 
condition~\cite{Weinberg:1962hj,Efimov:1993ei,Anikin:1995cf,%
Dong:2009tg, Faessler:2007gv}. It implies that the renormalization
constant of the hadron state is set equal to zero with
\eq
\label{ZLc} Z_H^I  = 
1 - \Big(\Sigma_H^I(p^2)\Big)^\prime\biggl|_{p^2 = M_H^2} = 0 \,.
\en
Here, $(\Sigma_H^I)^\prime$ is the derivative of the transverse part of
the mass operator $(\Sigma_H^I)^{\mu\nu}$ induced by the hadronic $(I=M)$ 
and the charm quark loop $(I=C)$, respectively. The mass operator including
transverse and longitudinal parts is defined as
\eq
\Sigma_H^{\mu\nu}(p) = g^{\mu\nu}_\perp \, \Sigma_H(p) +
\frac{p^\mu p^\nu}{p^2} \Sigma_H^L(p)\,, \quad g^{\mu\nu}_\perp =
g^{\mu\nu} - \frac{p^\mu p^\nu}{p^2} \,.
\en 
The diagrams corresponding to the 
contribution of the meson $(\Sigma_H^M)^{\mu\nu}$ and 
charm quark loops $(\Sigma_H^C)^{\mu\nu}$ are shown in Figs.~1 and~2. 
In the evaluation of meson and quark loop diagrams we use free 
propagators for the mesons and the charm quark with a mass 
$m_c = 2.16$ GeV~\cite{Anikin:1995cf}. In particular, 
the propagators for vector $D^\ast$ and axial $D$ mesons denoted by 
$D^\ast_{\alpha\alpha'}$ and $D_{1\,, \beta\beta'}$ are given by 
\eq   
{\cal D}_{\alpha\beta}(k) = \frac{1}{M_{\cal D}^2 - k^2} \,
\biggl[ - g_{\alpha\beta} + \frac{k_\alpha k_\beta}{k^2} \biggr]\,,
\quad {\cal D} = D^\ast, D_1
\en
and similarly for the pseudoscalar $D$ meson with
\eq
D(k) = \frac{1}{M_D^2 -k^2} \,.
\en
The propagator of the charm quark is written as 
\eq 
S_c(k) = \frac{1}{m_c-\!\not k} \,.
\en 

\begin{figure}[ht]
\centering\includegraphics[scale=0.55]{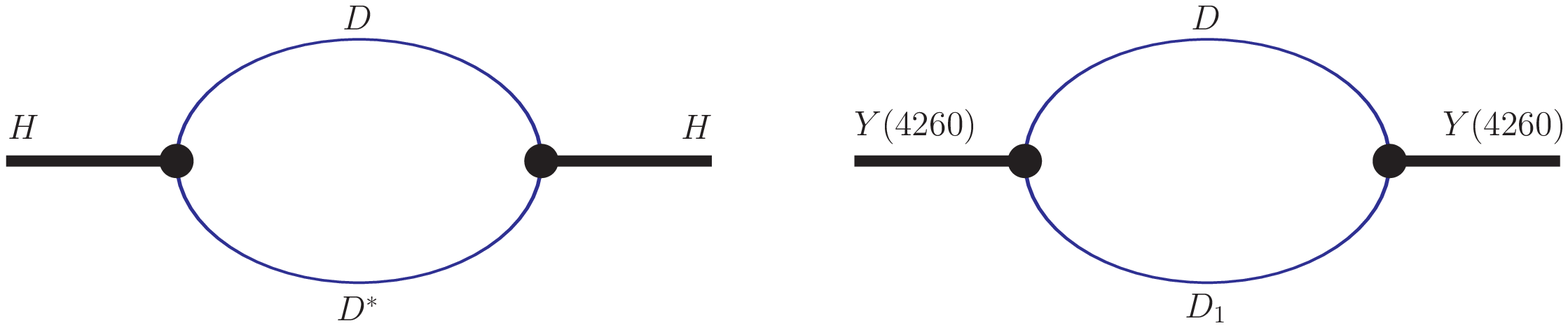}
\caption{Hadronic molecular component contributions 
to the mass operators of $X(3872)$ and $Y(4260)$.} 

\vspace*{1cm}
\centering\includegraphics[scale=0.55]{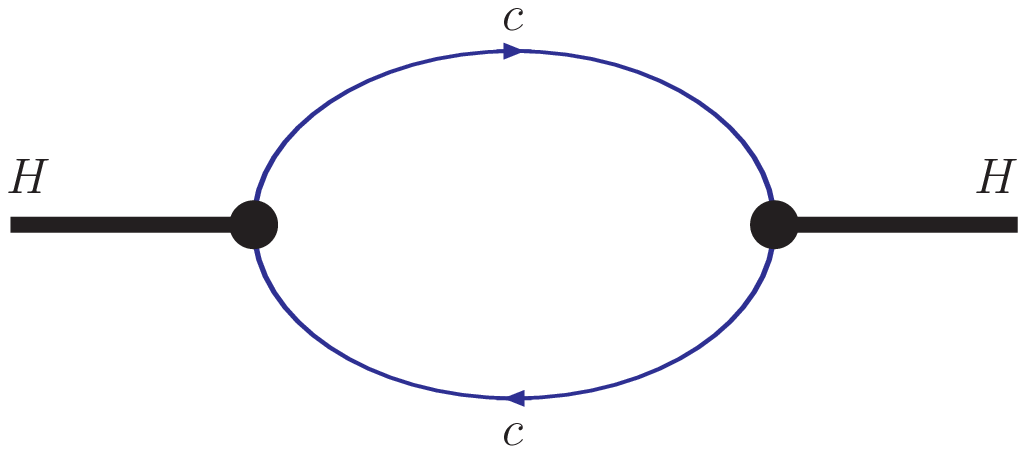}
\caption{Charmonium component contribution to the mass operators 
of $H = X(3872), Y(4260)$.} 
\end{figure}

The explicit expressions for the coupling constants $g_H^M$ 
and $g_H^C$ 
resulting from the compositeness condition are
\eq
(g_X^M)^{-2}&=& I(M_X^2,M_{D^0}^2,M_{D^{\ast 0}}^2) \,, \nonumber\\
(g_Y^M)^{-2}&=& \frac{1}{2} \, \biggl[ I(M_Y^2,M_{D^+}^2,M_{D^-_1}^2)
\,+\, I(M_Y^2,M_{D^0}^2,M_{D^0_1}^2)\biggr] 
\en 
and 
\eq
(g_X^C)^{-2} = J_-(M_X^2,m_c^2)\,, \quad 
(g_Y^C)^{-2} = J_+(M_Y^2,m_c^2)\, . 
\en 
The structure integrals $I$ and $J_\pm$ are defined as 
\eq
I(M_{H}^2,M_1^2,M_2^2) = \frac{M_H^2}{32\pi^2} \, \int\limits_0^\infty
\frac{d\alpha_1 d\alpha_2}{(t_H^M)^3} \, R_H(\alpha_1,\alpha_2)
\, \biggl( 1+\frac{1}{2 M_2^2 t_H} \biggr) \, \exp\Big[ - \alpha_1
M_1^2 - \alpha_2 M_2^2 + R_H(\alpha_1,\alpha_2) M_H^2\Big] 
\en
and 
\eq 
J_\pm(M_H^2,m_c^2) &=& \frac{3}{4\pi^2} \, \int\limits_0^\infty
\frac{d\alpha_1 d\alpha_2}{(t_H^C)^2} \,
\biggl( Q_H(\alpha_1,\alpha_2) + 
P_H(\alpha_1,\alpha_2) \biggl[ 
\pm m_c^2 + \frac{1}{t_H^C} + M_H^2 Q_H(\alpha_1,\alpha_2) \biggr] 
\biggr) \nonumber\\
&\times&\exp\Big[ 
- \alpha_{12} m_c^2 + P_H(\alpha_1,\alpha_2) M_H^2\Big] \,, 
\en 
where
\eq
& &t_H^I =  2s_H^I + \alpha_{12}\,, \quad \alpha_{12} = \alpha_1 +
\alpha_2\,, \quad s_H^I = \frac{1}{(\Lambda_H^I)^2}\,, \nonumber\\
& &R_H(\alpha_1,\alpha_2) = 2 \alpha_1 \alpha_2 + s_H^M
\alpha_{12} \,, \nonumber\\
& &P_H(\alpha_1,\alpha_2) = \frac{1}{4} \biggl( 
\alpha_{12} - \frac{(\alpha_1-\alpha_2)^2}{t_H^C} \biggr)  \,, \nonumber\\
& &Q_H(\alpha_1,\alpha_2) = \frac{1}{4} \biggl( 
1 - \frac{(\alpha_1-\alpha_2)^2}{(t_H^C)^2} \biggr)\, .  
\en 
Once the masses of the composite states are fixed, the values for
the corresponding couplings to their constituents can be extracted
from the compositeness condition. In this work we only consider
the neutral meson pair for the $X(3872)$ and both the neutral and
charged pairs for the $Y(4260)$. Values for the hadronic molecular 
couplings to the $X(3872)$ at the preferred value of cutoff parameter 
$\Lambda_X = 0.5$ GeV are listed in Table I. 
The coupling $g_X^M$ has a weak dependence 
on the binding energy $\epsilon_{X}$, 
which is in agreement with the scaling law of hadronic molecules 
found for example in Ref.~\cite{Branz:2008cb}: 
$g_{H}^M \sim \epsilon_{H}^{1/4}$. 
Note that the results for the coupling $g_Y^M$ at $\Lambda_Y^M = 0.75$ 
GeV is $g_Y^M = 5.72$. 
It increases to 6.13 when $\Lambda_Y^M$ decreases to  $0.5$ GeV.  
The charmonium coupling $g_X^C$ for different values of 
$\epsilon_X = 0.05 - 0.3$ MeV is very stable 
and changes from 9.287 at $\epsilon_X = 0.05$ MeV to 
9.289 at $\epsilon_X = 0.3$ MeV. 
The result for $g_Y^C$ is $g_Y^C = 2.03$. 

\begin{center}
{\bf Table I.} Results for the coupling constant $g_{_{X}}^M$
depending on $\epsilon_{X}$. 

\vspace*{.25cm} 
\def\arraystretch{1.5}
\begin{tabular}{|c|c|c|c|c|c|}\hline
\multicolumn{6}{|c|}{$\epsilon_X$ in MeV}\\[1mm]
\hline
0.05 & 0.1   & 0.15 & 0.2   & 0.25   & 0.3  \\ \hline
2.06 & 2.07  & 2.08 & 2.09  & 2.10   & 2.11 \\ \hline
\end{tabular}
\end{center} 

The diagrams contributing to the two-body radiative decays
$Y(4260)\to X(3872) + \gamma$ are summarized in Figs.~3 and~4. 
To calculate the radiative decay we need to specify how we couple the 
photons both for the hadronic molecular and the charmonium components. 

\begin{figure}
\centering \includegraphics[scale=0.75]{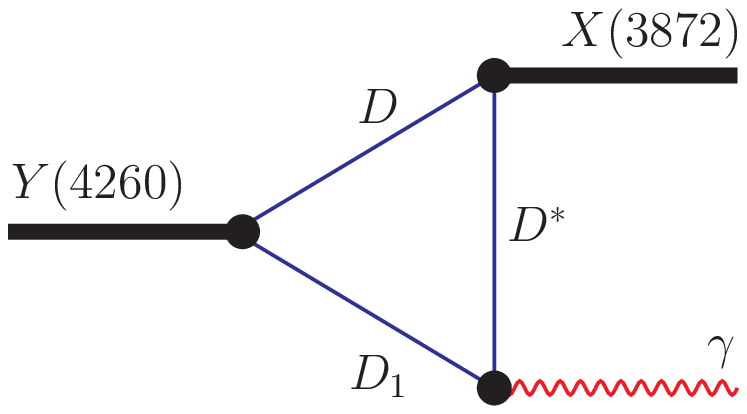}
\caption{Diagram describing the hadronic molecular component contribution 
to the radiative decay $Y(4260) \to X(3872) + \gamma$.}

\vspace*{1cm}
\centering \includegraphics[scale=0.75]{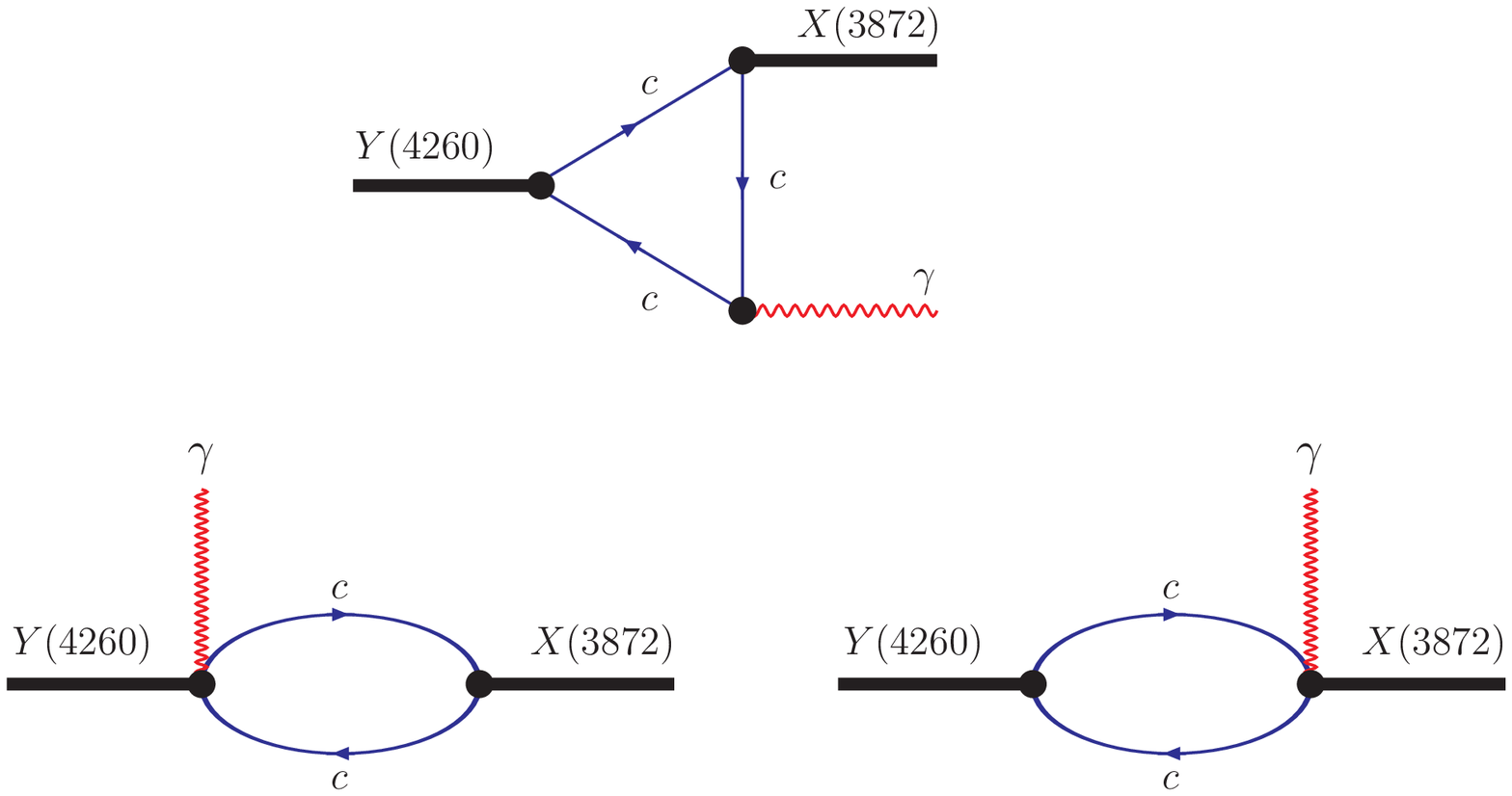}
\caption{Diagrams describing the charmonium  component contribution 
to the radiative decay $Y(4260) \to X(3872) + \gamma$.}
\end{figure}

For the hadronic case we need to specify the 
$D_1\to D^\ast\gamma$ coupling. 
The relevant phenomenological Lagrangian generating the
coupling  $D_1D^\ast\gamma$ is 
\eq\label{L_em_ext} 
{\cal L}_{D^\ast D_1 \gamma}(x) &=& \frac{e}{2}
\, F_{\mu\nu}(x) \, \epsilon^{\mu\nu\alpha\beta} \, \, \biggl(
  g_{_{D^{\ast \, \pm } D ^\pm_1 \gamma}}
 \, D^{\ast \, +}_{\alpha}(x) \, D_{1, \beta}^-(x)
+ g_{_{D^{\ast \, 0} D^0_1 \gamma}}
 \, D^{\ast \, 0}_{\alpha}(x) \, \bar D_{1, \beta}^0(x) \biggr)
 \, + \, {\rm H.c.}  \,, 
\en
where the couplings 
\eq
g_{_{D^{\ast \, 0} D^0_1 \gamma}}\,, \quad 
g_{_{D^{\ast \, \pm } D^\pm_1 \gamma}} = 
\frac{1}{4} g_{_{D^{\ast \, 0} D^0_1 \gamma}}   
\en 
can be fixed using the estimate for the 
rate $\Gamma(D_1^0 \to D^{\ast \, 0} \gamma)$ given 
in Ref.~\cite{Korner:1992pz} 
\eq 
\Gamma(D_1^0 \to D^{\ast \, 0} \gamma) = \frac{\alpha}{24} \, 
g_{_{D^{\ast \, 0} D^0_1 \gamma}}^2 \, M_{D_1^0} \, (1-r^2)^3 \, 
\biggl(1 + \frac{1}{r^2} \biggr) \,, \quad \quad 
r = \frac{M_{_{D^{\ast \, 0}}}}{M_{_{D_1}}} \,.  
\en 
Using the numerical estimate of Ref.~\cite{Korner:1992pz}
with $\Gamma(D_1^0 \to D^{\ast \, 0} \gamma) = 245 \pm 18$ keV 
we get $g_{_{D^{\ast \, 0} D^0_1 \gamma}} = 2.10 \pm 0.08$. 
Note that the relation between the couplings 
$g_{_{D^{\ast \, \pm } D^\pm_1 \gamma}}$ 
and $g_{_{D^{\ast \, 0} D^0_1 \gamma}}$ has been deduced in the naive quark 
model [see Refs.~\cite{Faessler:2008vc,Dong:2008gb}]. In particular, 
the corresponding ratio 
$R = g_{_{D^{\ast \, 0} 
D^0_1 \gamma}}/g_{_{D^{\ast \, \pm } D^\pm_1 \gamma}}$ is 
expressed in terms of the electric charges of the valence quarks
of the neutral and charged $D$ mesons 
\eq 
R = \frac{e_c+e_u}{e_c+e_d} = 4 \,. 
\en 
In case of charmonium we include the photons by direct coupling 
to the $c$-quarks as 
\eq\label{em_triangle}
{\cal L}_{\bar cc\gamma}(x) = 
\frac{2}{3}e \, A_\mu(x) \bar c(x) \gamma^\mu c(x)
\en 
and in the interaction Lagrangian 
${\cal L}_{H\bar c c}(x) = g_H^C H(x) J_H^C(x)$ in order to fulfill 
electromagnetic gauge invariance of these nonlocal Lagrangians. 
The gauging proceeds in a way suggested and extensively used
in Refs.~\cite{Anikin:1995cf,Mandelstam:1962mi,Terning:1991yt}. 
To guarantee local electromagnetic gauge invariance of the strong
interaction Lagrangian one multiplies each quark and Dirac conjugated 
quark field in ${\cal L}_{H\bar c c}$ with a gauge field exponential 
$I(y,x) = \int_x^y dz_\mu A^\mu(z)$  
as
\eq
c(y) \to e^{-i\frac{2}{3}e I(y,x)} c(y)\,, \quad 
\bar c(y) \to \bar c(y) e^{i\frac{2}{3}e I(y,x)} \,. 
\en 
As a result the gauge-invariant (GI) nonlocal Lagrangians describing 
the coupling of the $X(3872)$ and $Y(4260)$ states to the charm quarks read 
\eq\label{gauging}
{\cal L}_{H \bar cc}^{GI}(x) &=& H(x)^{\mu} J_{H;\mu}^{C; GI}(x) \,, 
\en 
where $J_H^{C; GI}(x)$ is the GI interpolating charm-quark current 
with quantum numbers of the $H=X(3872)$ or $Y(4260)$ state: 
\eq\label{em_bubble} 
J_{X;\mu}^{C; GI}(x) &=& 
\int d^4y \Phi_{X}^C(y^2)\bar{c}(x+y/2) \, \gamma_{\mu}\gamma^5 \, 
e^{i\frac{2}{3}e I(x+y/2,x-y/2)} \, c(x-y/2)\,, \nonumber\\
J_{Y;\mu}^{C; GI}(x) &=& 
\int d^4y \Phi_{Y}^C(y^2)\bar{c}(x+y/2) \, \gamma_{\mu} \, 
e^{i\frac{2}{3}e I(x+y/2,x-y/2)} \, c(x-y/2)\,. 
\en 
The Lagrangian~(\ref{em_triangle}) generates the triangle diagram of Fig.~4 
(see upper panel), while the two additional (bubble) diagrams 
in the lower panel of 
Fig.~4 are generated by terms of the Lagrangian~(\ref{em_bubble}) obtained 
after expansion of the gauge exponential up to first order in $A^\mu$. 

The diagram of Fig.~3 is separately gauge invariant because  
the $D_1D^{\ast 0}\gamma$ interaction Lagrangian~(\ref{L_em_ext}) 
contains a contraction of the antisymmetric Levi-Cevit\'a tensor 
and the stress tensor of the electromagnetic field. 
The proof of gauge invariance of the 
set of diagrams in Fig.~4 is presented in Appendix~A. 

\section{Decay modes and results}

The two-body decay width for the transition $Y(4260)\to X(3872) +
\gamma$ is given by 
\eq
\label{Gamma2} \Gamma(Y \to X \gamma) \, = \,
\frac{P^\ast}{24\pi \, M_Y^2} \ \sum_{\rm pol} \Big|M_{\rm inv}\Big|^2
\en
where $P^\ast = (M_Y^2 - M_H^2)/(2M_Y)$ is the three-momentum of the
final states in the rest frame of the $Y(4260)$ and 
$M_{\rm inv}$ is the corresponding invariant matrix element. 

The contribution of the hadronic loop diagram to the 
invariant matrix element for the radiative transition is written
as
\eq
M_{\rm inv}^M = e \epsilon_{\mu\nu\alpha\beta} \,
\epsilon^{\ast\nu}_\gamma(q) \, \epsilon^{\ast}_{_{X\alpha'}}(p') \,
\epsilon_{_{Y\beta'}}(p) \, q^\mu  \, I^{\alpha\alpha'\beta\beta'}_M
\en
where $p, p', q$ and $\epsilon_{_{Y\beta'}}(p)$,
$\epsilon^{\ast}_{_{X\alpha'}}(p')$, $\epsilon^{\ast\nu}_\gamma(q)$ 
are the momenta and polarization 
vectors of $Y(4260)$, $X(3872)$ and the photon, respectively. The
loop integral $I^{\alpha\alpha'\beta\beta'}_M$ contributing
to the radiative decay of the $Y(4260)$ state is given as
\eq
I^{\alpha\alpha'\beta\beta'}_M = g_{\rm eff}^M \, \int
\frac{d^4k}{\pi^2 i} \,
\Phi_Y\biggl(-\Big[k+\frac{p}{2}\Big]^2\biggr) \,
\Phi_X\biggl(-\Big[k+\frac{p}{2}+\frac{q}{2}\Big]^2\biggr)
\, D^{\ast 0}_{\alpha\alpha'}(k+q) \, D_{1\,,
\beta\beta'}^0(k) \, D^0(k+p) \,,
\en
where
\eq
g_{\rm eff}^M  &=&  \frac{1}{16 \pi^2 \sqrt{2}} \, 
g_{_{Y}} \, g_{_{X}} \, g_{_{D^{\ast \, 0 } D^0 \gamma}} \, M_Y \, M_X \, 
\cos\theta_X \, \cos\theta_Y 
\en
is the effective coupling constant. 
Using the Feynman $\alpha$ parametrization and performing the
integration over the loop momentum we can expand the loop integral
$I^{\alpha\alpha'\beta\beta'}$ in terms of three independent Lorentz
structures for all involved particles on mass shell with $p^2=M_Y^2,
p^{\prime\, 2} = M_X^2, q^2 = 0$~\cite{Dong:2008gb}
\eq
I^{\alpha\alpha'\beta\beta'}_M = \frac{1}{M_Y^2} 
\, \biggl(g^{\alpha\alpha'} g^{\beta\beta'}
\, pq \, F_1 \, + \, g^{\alpha\alpha'} p^\beta  \, q^{\beta'} \, F_2
\, + \, g^{\beta\beta'} p^\alpha \, q^{\alpha'} \, F_3  \biggr)\,.
\en
The $F_i$ are relativistic form factors given by integrals over
Feynman parameters. Note that our full phenomenological
Lagrangian describing the radiative transition of the $Y(4260)$ state
has a simple structure reflected in the
corresponding matrix element for the $Y(4260) \to X(3872) + \gamma$
decay. It contains a smaller number of linearly independent
Lorentz structures as was found for example in the
tetraquark model~\cite{Dubnicka:2011mm}.
In Ref.~\cite{Dubnicka:2011mm} it was also shown
that in general the matrix element for the radiative transition between axial
and vector meson states contains four independent Lorentz structures 
due to electromagnetic gauge invariance. However, as was also 
clearly demonstrated in Ref.~\cite{Dubnicka:2011mm} consideration of the 
additional Schouten identity, which forbids antisymmetric fifth-rank tensors 
in four dimensions [see details in Ref.~\cite{Korner:2003zq}], leads to a 
further simplification of the Lorentz structure of the 
$Y(4260) \to X(3872) + \gamma$ transition matrix element. 
The form factors $F_1$, $F_2$ and $F_3$ form two linearly independent 
combinations $F_1 + F_2$ and $F_1 + F_3$ corresponding to 
the $E_1$ and $M_2$ transitions~\cite{Dubnicka:2011mm}. 

The expressions for the form factors $F_i  $ are given by
\eq
F_i &=& g_{\rm eff}^M R_i(M_{D^0}^2,M_{D^{\ast 0}}^2) \,,
\en
where $R_i(M_1^2,M_2^2)$ are the structure integrals with
\eq
R_i(M_1^2,M_2^2) &=& \int\limits_0^\infty \frac{d\alpha_1
d\alpha_2 d\alpha_3}{\Delta^2} \, f_i \,
\exp[z(\alpha_1,\alpha_2,\alpha_3)] \,, \nonumber\\
f_1 &=& \biggl[ 1 + \frac{1}{2 \Delta} 
\biggl( \frac{1}{M_{D_1}^2} + \frac{1}{M_2^2} \biggr)\biggr] \, 
\frac{2 M_Y^2}{M_Y^2-M_X^2}\,, \nonumber\\
f_2 &=& - x_1 x_2 \frac{M_Y^2}{M_{D_1}^2} \,, \nonumber\\
f_3 &=& - x_1 (x_1 + x_2 - 1) \frac{M_Y^2}{M_2^2} \,,
\en
and
\eq
z(\alpha_1,\alpha_2,\alpha_3) &=& - \alpha_1 M_1^2 - \alpha_2
M_2^2 - \alpha_3 M_{D_1}^2
\nonumber\\
&+&  M_Y^2 \biggl(\alpha_1 + \frac{s_Y+s_X}{4} - x_1^2 \Delta\biggr)
\, + \,  (M_Y^2 - M_X^2) \biggl(\frac{s_X}{4} - x_1 x_2
\Delta\biggr)
\,, \nonumber\\
\Delta &=& \alpha_{123} + s_Y + s_X\,, \quad s_Y  =
\frac{1}{\Lambda_Y^2}\,, \ \
\alpha_{123} = \alpha_1 + \alpha_2 + \alpha_3 \,, \nonumber\\
x_1 &=& \frac{1}{\Delta} \bigg( \alpha_1 + \frac{s_Y}{2} +
\frac{s_X}{2} \biggr)\,, \ \ x_2 \ = \ \frac{1}{\Delta} \bigg(
\alpha_2 + \frac{s_X}{2} \biggr)\,.
\en
Finally, the matrix element of the radiative transition $Y(4260) \to
X(3872) + \gamma$ can be written in a manifestly gauge-invariant form
\eq
M_{YX\gamma}^M = \frac{e}{M_Y^2}
\biggl( \, 
\epsilon_{p'q\epsilon_\gamma^\ast\epsilon_{_{X}}^\ast} \, 
(\epsilon_{_{Y}} p') \, (F_1 + F_2) \, + \,  
\epsilon_{p'q\epsilon_\gamma^\ast\epsilon_{_{Y}}} \, 
(\epsilon_{_{X}}^\ast q) \, (F_1 + F_3) \, 
\biggr) \,,
\en 
where $\epsilon_{_{ABCD}} = A^\alpha B^\beta C^\rho D^\sigma 
\, \epsilon_{\alpha\beta\rho\sigma}$. 

The contributions of the charm quark-loop diagrams of Fig.~4 are given in 
Appendix~A. The corresponding matrix element is calculated using 
FORM~\cite{Vermaseren:2000nd} and is decomposed into four 
relativistic form factors factors $G_i$ $(i=1\ldots 4)$ 
as 
\eq
M^C_{YX\gamma} &=& \frac{e}{M_Y^2} \, 
\biggl( G_1 \Big[ (p'\epsilon_\gamma^\ast) \, 
\epsilon_{p'q\epsilon_{_{Y}}\epsilon_{_{X}}^\ast} 
- (p'q) \,  \epsilon_{p\epsilon_{_{Y}}\epsilon_{_{X}}^\ast\epsilon_\gamma^\ast}
            \Big] \,+\, 
        G_2 \, (q\epsilon_{_{X}}^\ast) \, 
        \epsilon_{p'q\epsilon_{_{Y}}\epsilon_\gamma^\ast} \nonumber\\
        &+&  
        G_3 \, (q \epsilon_{_{Y}}) \, 
        \epsilon_{p'q\epsilon_{_{X}}^\ast\epsilon_\gamma^\ast}  \,+\, 
        G_4 \, (p'q) \, 
\epsilon_{q\epsilon_{_{Y}}\epsilon_{_{X}}^\ast\epsilon_\gamma} 
\biggr) \, .
\en 
Again in full consistency with the proof of Ref.~\cite{Dubnicka:2011mm},    
the combination of the four form factors $G_i$ is reduced 
to a matrix element with two Lorentz structures, 
where the form factors $G_i$ form two linearly independent combinations,  
\eq 
M_{YX\gamma}^C &=& \frac{e}{M_Y^2}
\biggl( \, 
\epsilon_{p'q\epsilon_\gamma^\ast\epsilon_{_{X}}^\ast} \, 
(\epsilon_{_{Y}} p') \, 
\biggl(G_1 \frac{2M_Y^2}{M_Y^2-M_X^2} 
\,+\, G_3 \,-\, G_4\biggr) \nonumber\\
&+&\epsilon_{p'q\epsilon_\gamma^\ast\epsilon_{_{Y}}} \, 
(\epsilon_{_{X}}^\ast q) \, 
\biggl(G_1 \frac{M_Y^2+M_X^2}{M_Y^2-M_X^2} 
\,-\, G_2 \,-\, G_4\biggr) 
\biggr) \,. 
\en 
Combining the contributions of both the hadronic and charmonium components 
to the radiative transition $Y(4260) \to X(3872) + \gamma$ we form 
two linear-independent combinations of the corresponding form factors 
${\cal F}_1$ and ${\cal F}_2$:  
\eq 
{\cal F}_1 = F_1 + F_2 + G_1 \frac{2M_Y^2}{M_Y^2-M_X^2} 
\,+\, G_3 \,-\, G_4\,, \nonumber\\
{\cal F}_2 = F_1 + F_3 + G_1 \frac{M_Y^2+M_X^2}{M_Y^2-M_X^2} 
\,-\, G_2 \,-\, G_4\,. 
\en 
Then the decay width of the radiative transition 
$Y(4260) \to X(3872) + \gamma$ is expressed in terms 
of ${\cal F}_1$ and ${\cal F}_2$ as 
\eq
\Gamma(Y \to X + \gamma) \, = \,
\frac{\alpha}{3} \, \frac{(P^\ast)^5}{M_Y^4} \, \biggl[ 
{\cal F}_1^2 \, + \, \frac{M_Y^2}{M_X^2} {\cal F}_2^2 \biggr]
\,. 
\en

Numerical results for the radiative decay $\Gamma(Y(4260) \to
X(3872) + \gamma$ in a pure molecular scenario, 
which are about 50 keV, are
summarized in Table II. In our calculations we use different 
values of the binding energy 
$\epsilon_X$ = 0.05, 0.1, 0.15, 0.2, 0.25  and 0.3 MeV. 
When we also vary the cutoff parameters $\Lambda_X$ and $\Lambda_Y$ 
in the region $0.5-0.75$ GeV we finally get the following estimate for 
the decay rate in the molecular picture 
\eq 
\Gamma(Y \to X + \gamma)  = 59 \pm 12.4 \, \mathrm{keV} \,. 
\en 
Note that in Ref.~\cite{Guo:2013zbw} the result for 
$\Gamma(Y \to X + \gamma)$ is written in terms of the coupling 
constants $x$ and $c_0$ corresponding to the 
$XD^0D^{\ast 0}$ and $D_1^0D^{\ast 0}$ 
couplings, respectively~\cite{Guo:2013zbw} 
\eq 
\Gamma(Y \to X + \gamma)  = 141^{+136}_{-91} \, (x^2 \, \mathrm{GeV}) \, 
c_0^2 \, \mathrm{keV} \,, 
\en 
where in the numerical evaluation the coupling $c_0$ is taken as 0.4 
and the coupling $x$ is fixed as 
\eq 
|x| = 0.97^{+0.40}_{-0.97} \pm 0.14 \, \mathrm{GeV}^{-1/2} \,. 
\en 

\begin{center}
{\bf Table II.} Decay width for $Y(4260) \to X(3872) + \gamma$ in keV 
in a pure molecular scenario. 

\vspace*{.25cm}
\def\arraystretch{1.5}
\begin{tabular}{|c|c|c|c|c|c|}\hline
\multicolumn{6}{|c|}{$\epsilon_X$ in MeV}\\[1mm]
\hline 
0.05    & 0.10  &  0.15    & 0.20    & 0.25  & 0.30                 \\ \hline
                                       $52.4 \pm 4.0$         
                                     & $52.5 \pm 4.0$         
                                     & $52.6 \pm 4.1$         
                                     & $52.7 \pm 4.1$         
                                     & $52.9 \pm 4.1$         
                                     & $53.0 \pm 4.2$ 
      \\ \hline
\end{tabular}
\end{center}

Inclusion of the charmonium components in the $X(3872)$ 
and $Y(4260)$ states adds the dependence on the mixing angles 
$\theta_X$ and $\theta_Y$. In this case we restrict 
to the binding energy $\epsilon_X = 0.1$ MeV and the central value 
of $g_{_{D^{\ast \, 0} D^0_1 \gamma}} = 2.10$. 
The results for 
$\Gamma(Y \to H + \gamma)$ in terms of the mixing angles 
$\theta_X$ and $\theta_Y$ are given by  
\eq 
\Gamma(Y \to H + \gamma) = 52.5 \ {\rm keV} \, 
\biggl[ 0.42   (\cos\theta_X \cos\theta_Y + 1.51 \sin\theta_X\sin\theta_Y)^2 
\,+\, 0.58 (\cos\theta_X \cos\theta_Y + 1.23 \sin\theta_X\sin\theta_Y)^2 
\biggr]\,. 
\en 
Here the first and second terms in square brackets represent 
the contributions of the ${\cal F}_1$ and ${\cal F}_2$ form factors, 
respectively. One can see that the pure charmonium scenario [when the 
molecular contributions in $X(3872)$ and $Y(4260)$) are neglected]  
gives the decay rate $\Gamma(Y \to H + \gamma) = 96.3$ keV. 
In Ref.~\cite{Dong:2009uf} we considered 
three scenarios of the molecular-charmonium mixing in $X(3872)$, 
which give results for the $X(3872)$ decay rates compatible with data. 
In particular, for $\epsilon_X = 0.1$ MeV these scenarios 
correspond to the following choice of the mixing angle $\theta_X$: 
scenarios I   ($\theta_X = 68.9^0$), 
II  ($\theta_X = - 12.6^0$) and III ($\theta_X = - 20.2^0$). 
Keeping $\theta_Y$ as a free parameter and substituting the $\theta_X$ 
in the specific scenario we get 
\eq 
\Gamma(Y \to H + \gamma) = 6.8  \ {\rm keV} \, 
\biggl[ 0.42   (\cos\theta_Y + 3.91 \sin\theta_Y)^2 
\,+\, 0.58 (\cos\theta_Y + 3.19 \sin\theta_Y)^2 
\biggr]
\en 
for scenario I, 
\eq 
\Gamma(Y \to H + \gamma) = 50.0 \ {\rm keV} \, 
\biggl[ 0.42   (\cos\theta_Y - 0.34 \sin\theta_Y)^2 
\,+\, 0.58 (\cos\theta_Y - 0.27 \sin\theta_Y)^2 
\biggr]
\en 
for scenario II,  
\eq
\Gamma(Y \to H + \gamma) = 46.2 \ {\rm keV} \, 
\biggl[ 0.42   (\cos\theta_Y - 0.56 \sin\theta_Y)^2 
\,+\, 0.58 (\cos\theta_Y - 0.45 \sin\theta_Y)^2 
\biggr]
\en 
for scenario III. Therefore the radiative decay 
$Y(4260) \to X(3872) + \gamma$ can also serve as a tool 
for testing a possible molecular-charmonium mixture 
of the $Y(4260)$ state. 
For example considering the mixing angle $\theta_Y$ as a free 
parameter in Fig.~5 we predict the rate 
$\Gamma(Y(4260) \to X(3872) + \gamma)$ as a function 
of $\sin\theta_Y$ for all three scenarios of the mixing in the $X(3872)$.
The respective numerical values for $\Gamma(Y(4260) \to X(3872) + \gamma)$ 
are given in the range as 
\eq 
\Gamma(Y(4260)\to X(3872)+\gamma) =  48.6 \pm 41.8 \ \mathrm{keV}
\en 
in scenario I, 
\eq 
\Gamma(Y(4260)\to X(3872)+\gamma) =  25.3 \pm 24.7 \ \mathrm{keV}
\en 
for scenario II and 
\eq 
\Gamma(Y(4260)\to X(3872)+\gamma) =  23.2 \pm 23.0 \ \mathrm{keV}
\en 
for scenario III.  In the case of scenario I, where the charmonium
components dominates in the $X(3872)$, 
an increase of $\sin\theta_Y$ leads 
to an enhancement of the radiative decay rate. 
In the case of scenarios II and III, which rely on 
a dominant molecular component
in the $X(3872)$, we have the opposite behavior.
In the following we assume the point of view that 
the $Y(4260)$ state has a very small charmonium component 
(since presently no direct evidence for a large admixture exists). 
Then we have the following estimate for the radiative decay rate of
$\Gamma(Y(4260)\to X(3872)+\gamma) = 6.8$ keV (scenario I), 
50 keV (scenario II) and 46.2 (scenario III). 

\begin{figure}
\centering \includegraphics[scale=0.65]{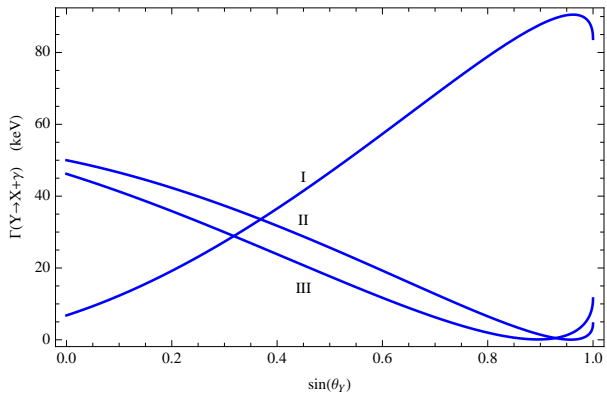}
\caption{Decay width $\Gamma(Y(4260) \to X(3872) + \gamma)$ 
as a function of the mixing angle $\theta_Y$ for 
three different scenarios of molecular-charmonium mixing in 
the $X(3872)$ state:  I   ($\theta_X = 68.9^0$), 
II  ($\theta_X = - 12.6^0$) and III ($\theta_X = - 20.2^0$).}
\end{figure}

In the recent BESIII~\cite{BESIII} 
measurement the product of the Born-order cross section
for $e^+e^- \to X(3872) \gamma$
times the branching fraction of $X(3872)\to J/\psi \pi^+\pi^-$
relative to the cross section of $e^+e^-\to J/\psi \pi^+\pi^-$
is determined as
\eq
\frac{\sigma^B[e^+e^- \to  X(3872) \gamma]\cdot {\cal B}[X(3872)\to
J/\psi\pi^+\pi^-]} {\sigma^B[e^+e^-\to J/\psi\pi^+\pi^-]}=(5.2\pm
1.9)\times 10^{-3}\;.
\en
The ratio is determined at the center-of-mass energy $\sqrt{s}=4.260$ GeV
under the assumption that the $X(3872)$ is produced only in $Y(4260)$
radiative decay. They also assume that the $e^+e^-\to X(3872) \gamma$
cross section follows that of $e^+e^-\to J/\psi\pi^+\pi^-$.
We further have the result of the Belle Collaboration~\cite{Belle} for
the branching fraction with
\eq
2.3\%<{\cal B}[X(3872)\to J/\psi\pi^+\pi^-]<6.6\%.
\en
For this range of values we can deduce the ratio 
\eq 
{\cal R}=\frac{\sigma^B[e^+e^-\to X(3872) \gamma]}
{\sigma^B[e^+e^-\to J/\psi\pi^+\pi^-]} = 11.7 \pm 7.1 \%.
\en
In Ref.~\cite{Dong:2013kta} we have analyzed the strong two- and
three-body decays of the $Y(4260)$ and found that 
\eq 
\Gamma(Y(4260) \to J/\psi \pi^+ \pi^-) = 1 \pm 0.2 \ {\rm MeV} \,. 
\en  
Using this prediction we obtain ratios of decay rates with
\eq
R^I_Y=\frac{\Gamma(Y(4260)\to X(3872) \gamma)} {\Gamma(Y(4260)\to
\pi^+\pi^-J/\Psi)}= 4.89 \pm 4.29 \% 
\en
for the scenario I, 
\eq
R^{II}_Y=\frac{\Gamma(Y(4260)\to X(3872) \gamma)} {\Gamma(Y(4260)\to
\pi^+\pi^-J/\Psi)}= 2.50 \pm 2.49 \% 
\en
for II, and 
\eq
R^{III}_Y=\frac{\Gamma(Y(4260)\to X(3872) \gamma)} {\Gamma(Y(4260)\to
\pi^+\pi^-J/\Psi)}= 2.31^{+2.34}_{-2.31} \%
\en
for the scheme III.
All three values (within errors) are close to the lower 
range of the present experimental result for ${\cal R}$.

In summary, using a phenomenological Lagrangian approach we give
predictions for the radiative two-body decay rate $Y(4260)\to
X(3872)+\gamma$. Both for the $X(3872)$ and the $Y(4260)$
we also include besides the molecular a charmonium component.
In the case of the $X(3872)$ mixing schemes were previously fixed
by the known strong and radiative decays. The full prediction
for the radiative decay rate $Y(4260)\to X(3872)+\gamma$ is contained in 
Fig.~5 and spans the values from 0 to about 85 keV, depending on the 
charmonium admixture in the $Y(4260)$ and the dominant component in the 
$X(3872)$. For a $X(3872)$ which has leading molecular component and 
a $Y(4260)$ which has a vanishing charmonium content the prediction for 
the radiative rate is about 35 keV. This result together with the one for
the strong three-body decay 
mode is able to explain the measured ratio ${\cal R}$ within errors
for this interpretation of the $Y(4260)$. 
It obviously would be helpful to also have a direct measurement of
the radiative $Y(4260)\to X(3872)+\gamma$ mode to test the structure
scenario given for the $Y(4260)$ and the $X(3872)$.

\begin{acknowledgments}

This work is supported by the DFG under Contract No. LY 114/2-1, 
by Tomsk State University Competitiveness Improvement Program, 
National Sciences Foundations of China Grants No. 10975146 and No. 11035006, 
and by the DFG and the NSFC through funds provided to the sino-German
CRC 110 ``Symmetries and the Emergence of Structure in QCD''. 
Y.B.D. thanks the Institute of Theoretical
Physics, University of T\"ubingen for the warm hospitality and is 
thankful for the support from the Alexander von Humboldt Foundation.

\end{acknowledgments}

\appendix\section{Gauge-invariance of diagrams in Fig.~4.}  

Here, we prove the gauge invariance of the set of diagrams of Fig.~4.  
The sum of the
contribution $M_{\rm inv}^{C; \triangle}$ of the triangle diagram 
generated by the direct coupling of charm quarks to the electromagnetic 
field and of the two bubble diagrams $M_{\rm inv}^{C; {\rm bub}}$  
generated by gauging the nonlocal strong Lagrangian 
describing the coupling of $X$ and $Y$ states to constituent 
charm quarks is denoted by the invariant matrix element $M_{\rm inv}^C$.
It is given by
\eq
M_{\rm inv}^C = M_{\rm inv}^{C; \triangle} + M_{\rm inv}^{C; {\rm bub}} \,, 
\en 
where 
\eq 
M_{\rm inv}^{C; i} = e \, \epsilon^{\ast\nu}_\gamma(q) \, 
\epsilon^{\ast}_{_{X\alpha}}(p') \,
\epsilon_{_{Y\beta}}(p) \, I_{\alpha\beta\nu}^{C; i}\,. 
\en
$I_{\alpha\beta\nu}^{C; i}$ are structure integrals 
generated by the triangle $i = \triangle$ and bubble $i = {\rm bub}$ 
diagrams 
\eq
I_{\alpha\beta\nu}^{C; \triangle} &=& - g_{\rm eff}^C 
\, \int
\frac{d^4k}{\pi^2 i} \,
\Phi_Y^C(-k^2) \,
\Phi_X^C(-(k+q/2)^2) \,
{\rm tr}\biggl( \gamma^{\alpha}\gamma^5 
S_c(k+p/2) \gamma^{\beta} 
S_c(k-p/2) \gamma^{\nu} 
S_c(k+p/2-p')
\biggr) \,,\nonumber\\
& &\\
I_{\alpha\beta\nu}^{C; {\rm bub}} &=& - g_{\rm eff}^C 
\, \int
\frac{d^4k}{\pi^2 i} \,
\biggl(\Phi_Y^C(-[k^2-kqt+q^2t/4])\biggr)^\prime \, 
\Phi_X^C(-k^2) \, (k-q/4)^\nu \, 
{\rm tr}\biggl( \gamma^{\alpha}\gamma^5 
S_c(k+p'/2) \gamma^{\beta} 
S_c(k-p'/2) \biggr) \nonumber\\
&-& g_{\rm eff}^C 
\, \int
\frac{d^4k}{\pi^2 i} \,
\Phi_Y^C(-k^2) \, 
\biggl(\Phi_X^C(-[k^2+kqt+q^2t/4])\biggr)^\prime \, 
\, (k+q/4)^\nu \, 
{\rm tr}\biggl( \gamma^{\alpha}\gamma^5 
S_c(k+p/2) \gamma^{\beta} 
S_c(k-p/2) \biggr) \,,\nonumber
\en
where 
\eq
g_{\rm eff}^C  &=&  \frac{g_{_{Y}}^C \, g_{_{X}}^C}{4 \pi^2} \,
\sin\theta_X \, 
\sin\theta_Y  
\en
is the effective coupling constant and 
\eq 
\biggl(\Phi_H^C(-z)\biggr)^\prime = 
\frac{\Phi_H^C(-z)}{dz} 
\en 
is the derivative of the correlation function. 

Contraction of the structure integral 
$I_{\alpha\beta\nu}^{C; \triangle}$ with the  photon momentum $q^\nu$ gives 
\eq 
q^\nu \, I_{\alpha\beta\nu}^{C; \triangle} &=& 
- g_{\rm eff}^C 
\, \int
\frac{d^4k}{\pi^2 i} \,
\Phi_Y^C(-k^2) \,
\Phi_X^C(-(k+q/2)^2) \nonumber\\
&\times& 
\biggl( {\rm tr}\biggl( \gamma^{\alpha}\gamma^5 
S_c(k+p/2) \gamma^{\beta} 
S_c(k+p/2-p')\biggr) 
\,-\, 
{\rm tr}\biggl( \gamma^{\alpha}\gamma^5 
S_c(k+p/2) \gamma^{\beta} 
S_c(k-p/2)\biggr) \biggr) \,. 
\en 
Here we used the set of Ward-Takahashi identities on the quark level:  
\eq
& &\not\! q = S_c^{-1}(k-p/2) - S_c^{-1}(k+p/2-p')  \,, \nonumber\\
& &S_c(k-p/2) \, \not\! q \, S_c(k+p/2-p') = 
S_c(k+p/2-p') - S_c(k-p/2) \,. 
\en 
The contraction of the 
$I_{\alpha\beta\nu}^{C; {\rm bub}}$ with the photon momentum $q^\nu$ 
gives
\eq 
q^\nu \, I_{\alpha\beta\nu}^{C; {\rm bub}} \equiv 
- q^\nu \, I_{\alpha\beta\nu}^{C; \triangle} + 
\Delta I_{\alpha\beta} 
\en  
where the term 
\eq 
\Delta I_{\alpha\beta} 
&=& g_{\rm eff}^C 
\, \int
\frac{d^4k}{\pi^2 i} \,
\Phi_Y^C(-k^2) \,
\Phi_X^C(-k^2) \, 
\biggl[ 
{\rm tr}\biggl( \gamma^{\alpha}\gamma^5 
S_c(k+p/2) \gamma^{\beta} 
S_c(k-p/2)\biggr) \nonumber\\
&-& 
{\rm tr}\biggl( \gamma^{\alpha}\gamma^5 
S_c(k+p'/2) \gamma^{\beta} 
S_c(k-p'/2)\biggr) \biggr] \equiv 0 
\en 
explicitly vanishes because of spinor traces. 

Therefore, the sum of triangle and bubble diagrams is manifestly 
gauge-invariant 
\eq 
q^\nu \, \biggl(
I_{\alpha\beta\nu}^{C; \triangle} \,+\, 
I_{\alpha\beta\nu}^{C; {\rm bub}} 
\biggr) \equiv 0 \,. 
\en


\begin{thebibliography}{99}

\bibitem{Faessler:2007gv}
  A.~Faessler, T.~Gutsche, V.~E.~Lyubovitskij and Y.~-L.~Ma,
  Phys.\ Rev.\ D {\bf 76}, 014005 (2007); 
  Phys.\ Rev.\ D {\bf 76}, 114008 (2007); 
  A.~Faessler, T.~Gutsche, S.~Kovalenko and V.~E.~Lyubovitskij,
  Phys.\ Rev.\ D {\bf 76}, 014003 (2007).

\bibitem{Faessler:2008vc}
  A.~Faessler, T.~Gutsche, V.~E.~Lyubovitskij and Y.~-L.~Ma,
  Phys.\ Rev.\ D {\bf 77}, 114013 (2008). 

\bibitem{Dong:2008gb}
  Y.~Dong, A.~Faessler, T.~Gutsche and V.~E.~Lyubovitskij,
  Phys.\ Rev.\ D {\bf 77}, 094013 (2008); 
  Y.~Dong, A.~Faessler, T.~Gutsche, S.~Kovalenko and V.~E.~Lyubovitskij,
  Phys.\ Rev.\ D {\bf 79}, 094013 (2009).

\bibitem{Branz:2008cb}
  T.~Branz, T.~Gutsche and V.~E.~Lyubovitskij,
  Phys.\ Rev.\ D {\bf 79}, 014035 (2009); 
  T.~Branz, T.~Gutsche and V.~E.~Lyubovitskij,
  Phys.\ Rev.\ D {\bf 80}, 054019 (2009);
  T.~Branz, T.~Gutsche and V.~E.~Lyubovitskij,
  Phys.\ Rev.\ D {\bf 82}, 054025 (2010).

\bibitem{Dong:2009uf}
  Y.~Dong, A.~Faessler, T.~Gutsche and V.~E.~Lyubovitskij,
 J.\ Phys.\ G {\bf 38}, 015001 (2011).

\bibitem{Dong:2009tg}
  Y.~Dong, A.~Faessler, T.~Gutsche and V.~E.~Lyubovitskij,
  Phys.\ Rev.\ D {\bf 81}, 014006 (2010);
  Y.~Dong, A.~Faessler, T.~Gutsche and V.~E.~Lyubovitskij,
  Phys.\ Rev.\ D {\bf 81}, 074011 (2010);
  Y.~Dong, A.~Faessler, T.~Gutsche, S.~Kumano and V.~E.~Lyubovitskij,
  Phys.\ Rev.\ D {\bf 82}, 034035 (2010);
  Phys.\ Rev.\ D {\bf 83}, 094005 (2011).


\bibitem{Dong:2012hc}
  Y.~Dong, A.~Faessler, T.~Gutsche and V.~E.~Lyubovitskij,
  J.\ Phys.\ G {\bf 40}, 015002 (2013).

\bibitem{Dong:2013iqa}
  Y.~Dong, A.~Faessler, T.~Gutsche and V.~E.~Lyubovitskij,
   Phys.\ Rev.\ D {\bf 88}, 014030 (2013).

\bibitem{Dong:2013kta}
  Y.~Dong, A.~Faessler, T.~Gutsche and V.~E.~Lyubovitskij,
  Phys.\ Rev.\ D {\bf 89}, 034018 (2014).

\bibitem{Weinberg:1962hj}
  S.~Weinberg,
  Phys.\ Rev.\  {\bf 130}, 776 (1963);
  A.~Salam,
  Nuovo Cim.\  {\bf 25}, 224 (1962);
  K.~Hayashi, M.~Hirayama, T.~Muta, N.~Seto and T.~Shirafuji,
  Fortsch.\ Phys.\ {\bf 15}, 625 (1967).

\bibitem{Efimov:1993ei}
  G.~V.~Efimov and M.~A.~Ivanov,
  {\it The Quark Confinement Model of Hadrons},
  (IOP Publishing, Bristol $\&$ Philadelphia, 1993).

\bibitem{Anikin:1995cf}
  I.~V.~Anikin, M.~A.~Ivanov, N.~B.~Kulimanova and V.~E.~Lyubovitskij,
  Z.\ Phys.\ C {\bf 65}, 681 (1995);
  M.~A.~Ivanov, M.~P.~Locher and V.~E.~Lyubovitskij,
  Few Body Syst.\  {\bf 21}, 131 (1996);
  M.~A.~Ivanov, V.~E.~Lyubovitskij, J.~G.~K\"orner and P.~Kroll,
  Phys.\ Rev.\ D {\bf 56}, 348 (1997);
  T.~Gutsche, M.~A.~Ivanov, J.~G.~Korner, V.~E.~Lyubovitskij and P.~Santorelli,
  Phys.\ Rev.\ D {\bf 86}, 074013 (2012);
  Phys.\ Rev.\ D {\bf 87}, 074031 (2013).

\bibitem{Barnes:2003vb} 
  T.~Barnes and S.~Godfrey, 
  Phys.\ Rev.\  D {\bf 69}, 054008 (2004).  
\bibitem{Barnes:2005pb}
  T.~Barnes, S.~Godfrey and E.~S.~Swanson,
  Phys.\ Rev.\  D {\bf 72}, 054026 (2005). 

\bibitem{Li:2009zu}
  B.~Q.~Li and K.~T.~Chao,
  Phys.\ Rev.\  D {\bf 79}, 094004 (2009).
\bibitem{Guo:2013zbw}
  F.~-K.~Guo, C.~Hanhart, U.~-G.~Mei{\ss}ner, Q.~Wang and Q.~Zhao,
  Phys.\ Lett.\ B {\bf 725}, 127 (2013).

\bibitem{Ma:2014ofa}
  L.~Ma, Z.~-F.~Sun, X.~-H.~Liu, W.~-Z.~Deng, X.~Liu and S.~-L.~Zhu,
  Phys.\ Rev.\ D {\bf 90}, 034020 (2014).

\bibitem{Zhu:2005hp}
  S.~-L.~Zhu,
  Phys.\ Lett.\ B {\bf 625}, 212 (2005). 

\bibitem{Ding:2008gr}
  G.~-J.~Ding,
  Phys.\ Rev.\ D {\bf 79}, 014001 (2009). 

\bibitem{Wang:2013cya}
  Q.~Wang, C.~Hanhart and Q.~Zhao,
  Phys.\ Rev.\ Lett.\  {\bf 111}, 132003 (2013). 

\bibitem{Swanson:2003tb} 
  E.~S.~Swanson,
  Phys.\ Lett.\ B {\bf 588}, 189 (2004). 

\bibitem{Korner:1992pz} 
  J.~G.~Korner, D.~Pirjol and K.~Schilcher,
  Phys.\ Rev.\ D {\bf 47}, 3955 (1993). 

\bibitem{Mandelstam:1962mi}
S.~Mandelstam,
Annals Phys.\  {\bf 19}, 1 (1962).

\bibitem{Terning:1991yt}
J.~Terning,
Phys.\ Rev.\ D {\bf 44}, 887 (1991).

\bibitem{Dubnicka:2011mm}
  S.~Dubnicka, A.~Z.~Dubnickova, M.~A.~Ivanov, J.~G.~Korner,
  P.~Santorelli and G.~G.~Saidullaeva,
 Phys.\ Rev.\ D {\bf 84}, 014006 (2011).

\bibitem{Korner:2003zq} 
  J.~G.~Korner and M.~C.~Mauser,
  Lect.\ Notes Phys.\  {\bf 647}, 212 (2004). 

\bibitem{Vermaseren:2000nd}
   J.~A.~M.~Vermaseren,
   Nucl.\ Phys.\ Proc.\ Suppl.\  {\bf 183}, 19 (2008);
   arXiv:math-ph/0010025.

\bibitem{BESIII}  M.~Ablikim {\it et al.} (BESIII Collaboration),
  Phys.\ Rev.\ Lett.\  {\bf 112}, 092001 (2014).  

\bibitem{Belle} C.~-Z.~Yuan (Belle Collaboration),
  arXiv:0910.3138 [hep-ex].

\end{thebibliography}
\end{document}